\documentclass[aps,prb,twocolumn,groupedaddress,superscriptaddress,amsfonts,amssymb,amsmath,floatfix,citeautoscript]{revtex4-1} 
\usepackage{graphicx} 
\usepackage[utf8]{inputenc}
\usepackage[centering,hmargin=25mm,tmargin=25mm,bmargin=25mm]{geometry}
\usepackage{amsmath}
\usepackage{hyperref} 
\usepackage{multirow}
\usepackage{newtxtext}
\usepackage{comment}
\usepackage{booktabs}
\usepackage{xcolor} 
\usepackage{soul}
\usepackage{hyperref}
\usepackage{float}
\usepackage[normalem]{ulem}
\hypersetup{colorlinks, 
    linkcolor={blue!75!black!80!yellow},
    citecolor={blue!75!black!80!yellow}, 
    urlcolor={blue!75!black!80!yellow}
    }
\usepackage[cmintegrals]{newtxmath}

\makeatletter
\renewcommand\@make@capt@title[2]{%
    \@ifx@empty\float@link{\@firstofone}{\expandafter\href\expandafter{\float@link}}%
    \sffamily{\textbf{#1}}\@caption@fignum@sep#2
}%

\makeatother

\usepackage{graphicx}
\usepackage{dcolumn}
\usepackage{bm}
\usepackage{hyperref}

\usepackage{csquotes}
\MakeOuterQuote{"}

\begin{document}

\title{Unboxing Quantum Black Box Models: Learning Non-Markovian Dynamics}

\author{Stefan Krastanov}
\affiliation{John A. Paulson School of Engineering and Applied Sciences, Harvard University, Cambridge, MA 02138, USA} 
\affiliation{Department of Electrical Engineering and Computer Science, Massachusetts Institute of Technology, Cambridge, MA 02139, USA}
\affiliation{Yale Quantum Institute, Yale University, New Haven, CT 06520, USA}

\author{Kade Head-Marsden}
\affiliation{John A. Paulson School of Engineering and Applied Sciences, Harvard University, Cambridge, MA 02138, USA} 

\author{Sisi Zhou}
\affiliation{
Department of Physics, Yale University, New Haven, Connecticut 06511, USA}
\affiliation{Pritzker School of Molecular Engineering, University of Chicago, Chicago, IL 60637, USA}

\author{Steven T. Flammia}
\affiliation{AWS Center for Quantum Computing, Pasadena, CA 91125, USA}

\author{Liang Jiang}
\affiliation{Yale Quantum Institute, Yale University, New Haven, CT 06520, USA}
\affiliation{Pritzker School of Molecular Engineering, University of Chicago, Chicago, IL 60637, USA}

\author{Prineha Narang}
\email{prineha@seas.harvard.edu}
\affiliation{John A. Paulson School of Engineering and Applied Sciences, Harvard University, Cambridge, MA 02138, USA} 

\date{\today}

\begin{abstract}



Characterizing the memory properties of the environment has become critical for the high-fidelity control of qubits and other advanced quantum systems. However, current non-Markovian tomography techniques are either limited to discrete superoperators, or they employ machine learning methods, neither of which provide physical insight into the dynamics of the quantum system. To circumvent this limitation, we design learning architectures that explicitly encode physical constraints like the properties of completely-positive trace-preserving maps in a differential form. This method preserves the versatility of the machine learning approach without sacrificing the efficiency and fidelity of traditional parameter estimation methods. Our approach provides the physical interpretability that machine learning and opaque superoperators lack. Moreover, it is aware of the underlying continuous dynamics typically disregarded by superoperator-based tomography. This paradigm paves the way to noise-aware optimal quantum control and opens a path to exploiting the bath as a control and error mitigation resource.
\end{abstract}

\maketitle

\section*{Introduction}

Understanding open quantum systems is critical to address and overcome the imperfections of current state-of-the-art quantum hardware. An array of master equation techniques have been developed to address the dynamics of both Markovian and non-Markovian open quantum systems from first principles~\citep{Breuer2002}. However, determining the appropriate forms and parameters for these master equation for real systems is challenging~\cite{Cubitt2012}. One way of linking these mathematical models to experiments is either through continuous parameter estimation or discrete operations tomography, which start with a model based in physical reality and fit the unknown parameters to experimentally available data. Considerable work has been dedicated to reconstruct master equations governing the dynamics of open quantum systems~\citep{Boulant2003, Perez2011, Bairey2020, Huang2020} including direct characterization~\citep{Mohseni2006, Mohseni2007} and the use of quantum process tomography for Lindblad estimations\citep{Chuang1997, tomography_poyatos1997complete, Howard2006,Bongioanni2010, Modi2012, Knee2018}. Moreover, some recent work has been dedicated to calibrating hardware beyond the Lindbladian formalism to characterize non-Markovian processes~\citep{Modi2012, Chen2020, Banchi2018, Bennink2019, Luchnikov2020, niu2019learning}.

To overcome the limitations of some of these methods, the physics community has started incorporating~\citep{minkov2020inverse,leung2017speedup,krastanov2019stochastic, Carleo2019} modeling, optimization, and gradient descent tools from machine learning, mainly in the form of advanced gradient descent algorithms~\citep{kingma2014adam} and automatic differentiation compilers~\citep{abadi2016tensorflow,paszke2017automatic,innes2018don}. Such tools were already in some limited use before the latest artificial intelligence revival, under the names of \emph{sensitivity analysis}, \emph{reverse design}, and even \emph{optimal control}~\citep{controlable_peirce1988optimal, compilation_dawson2005solovay, krotov_krotov1983iterative, grape_khaneja2005optimal, crab_caneva2011chopped}. However, typical parameter estimation methods fail when we cannot construct a model to be parameterized and fitted, causing a structural or bias error in the estimator (e.g. when a Lindbladian master equation fails to describe non-Markovian dynamics). In recent years deep learning has been used to circumvent this problem, formalized under the umbrella of the bias-variance theorem~\citep{schmidhuber2015deep,neal2018modern,belkin2019reconciling}. In principle these methods are capable of representing many useful dynamics that might be difficult to represent without bias errors when using typical approaches~\citep{flurin2020using, Luchnikov2020,carrasquilla2017machine}. However, such black box approaches swing far in the opposite direction of the trade-off imposed by the bias-variance theorem. Given their universality, deep learning constructs are not susceptible to the bias-inducing model errors that crop up when we use an incomplete description for the system under study, but they have vastly larger variance and require large amounts of experimentally-derived training data to overcome it. There are also examples of protocols that dynamically explore this trade-off, at least for Hamiltonian models, by modifying the structure of the model on-the-fly~\citep{gentile2020learning}.

\begin{figure}
    \centering
    \includegraphics[width=\columnwidth]{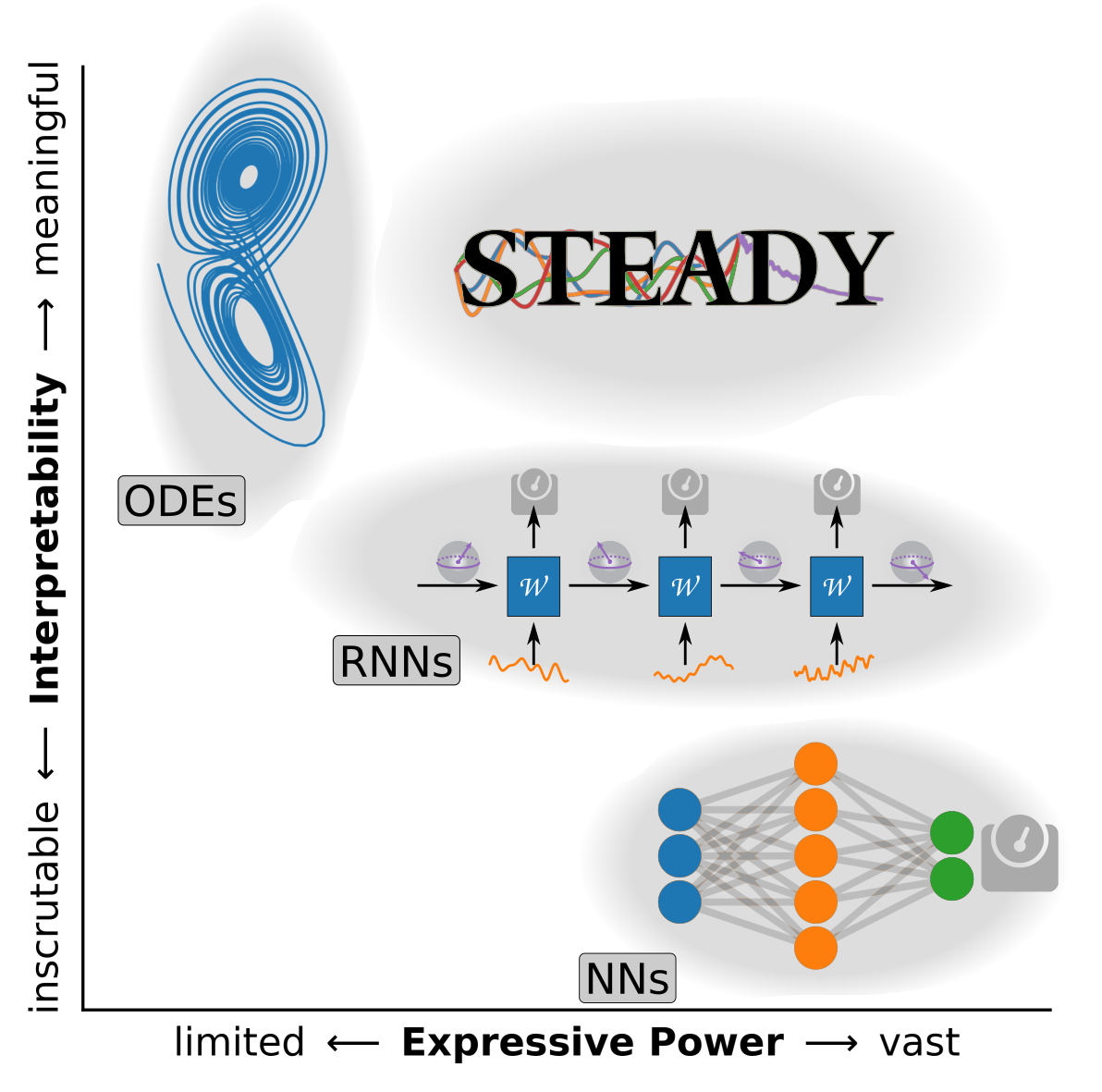}
    \caption{Susceptibility to model errors vs. efficiency and interpretability of the model: A richly parameterized model like a recurrent neural network need not suffer model errors, as long as vast amounts of training data is available. In contrast, a typical physics-informed model would have much lower variance for the same amount of training data, but might suffer from bias and model errors. The STEADY approach provides for richly over-parameterized modeling that inherits the useful properties of neural networks, without sacrificing as much interpretability and efficiency. STEADY is a new way to look at the bias-variance tradeoff.}
    \label{fig:interpretability_vs_richness}
\end{figure}

Despite the success of these black-box machine learning models, they have a significant ideological and practical weakness. A neural network might be sufficient to model and control a system, but it does not provide succinct insight into the underlying physics. More practically, a neural network approach also disregards the already available physical insight and does not exploit the simple, albeit imprecise, models we already have. Such `tabula rasa' approaches become particularly expensive, as the training process needs to learn the sum total of physics knowledge about the system from scratch, instead of starting from a nearly correct physical model.

Some of the physical constraints can be put back in the model, by throwing away any attempts to study the continuous dynamics, and simply learn the superoperator being realized by the given hardware. A large family of tomography techniques~\citep{tomography_leibfried1996experimental,tomography_chuang1997prescription,Kimmel2013,Knee2018,tomography_merkel2013selfconsistent,gstomography_greenbaum2015introduction} employ this approach, even for study of non-Markovian dynamics~\citep{Modi2012,Pollock2018,Pollock2018a,Bennink2019,Chen2020}. However, even this approach, does not provide insight into the continuous dynamics of the hardware, as only the final state of the dynamics is studied.

To fill this critical gap, here we present a continuous-dynamics modeling approach, named non-Markovian Stochastic Estimation of Dynamical Variables (STEADY). Our method requires comparatively little data to achieve a given level of variance, without sacrificing physicality. Moreover, our approach retains interpretability of the continuous dynamics, usually seen only in the explicit ODE modeling techniques, without suffering from their propensity for model errors. Figure~\ref{fig:interpretability_vs_richness} schematically depicts the trade-off between the versatility of machine learning (ML) models and the interpretability of traditional parameter estimation. ML avoids structural errors, as it is almost model-free. Traditional parameter estimation over master equations is interpretable, sparse, and less demanding, but susceptible to model errors. The STEADY method takes the best features of each approach, leading to a physical model based on known dynamical laws, but parameterized richly enough as to avoid the structural errors typically plaguing parameter estimation~\citep{krastanov2019stochastic}. Similarly to how convolutional networks exploit spatial structure~\citep{rawat2017deep} or recurrent networks encode memory~\citep{werbos1990backpropagation}, the architecture of our augmented model encodes the completely-positive trace-preserving (CPTP) nature of quantum dynamics, from the Markovian to the non-Markovian regimes. This enables the training procedure to proceed with comparatively little available training data, without sacrificing model fidelity. Importantly, our method learns the generators of the continuous dynamics of the system, not only discrete gates as done in many tomographic techniques~\citep{Bennink2019,greenbaum2015introduction}. This enables its use in optimal control and experimental design, where targets like operational fidelity and Fisher information are being optimized.

Next, we describe the architecture of the protocol and show how it compares with machine learning approaches. A particularly pertinent example of non-Markovian dynamics which have challenged conventional characterization methods comes from the quest to implement a reliable physical qubit. Whether in the case of solid state silicon~\citep{Loss1998, Kane1998, Breuer2004}, NV-centers~\cite{Haase2018, Dong2018}, or superconducting qubits~\cite{niu2019learning}, couplings to the bath of parasitic two-level systems cause significant non-Markovian effects. To retain this practical context, first we demonstrate the much higher efficiency of our method in characterizing a qubit undergoing weak measurements, when compared to recent machine learning techniques~\citep{gambetta2008quantum,flurin2018rnn}. Going beyond stochastic master equations, we then test our approach in characterizing a non-Markovian environment, a task previously attempted only through perturbative expansions of the Nakajima–Zwanzig equation~\citep{Nakajima1958, Zwanzig1960, Breuer2002}.

\section*{Results}

\begin{figure}
    \centering
    \includegraphics[width=0.45\textwidth]{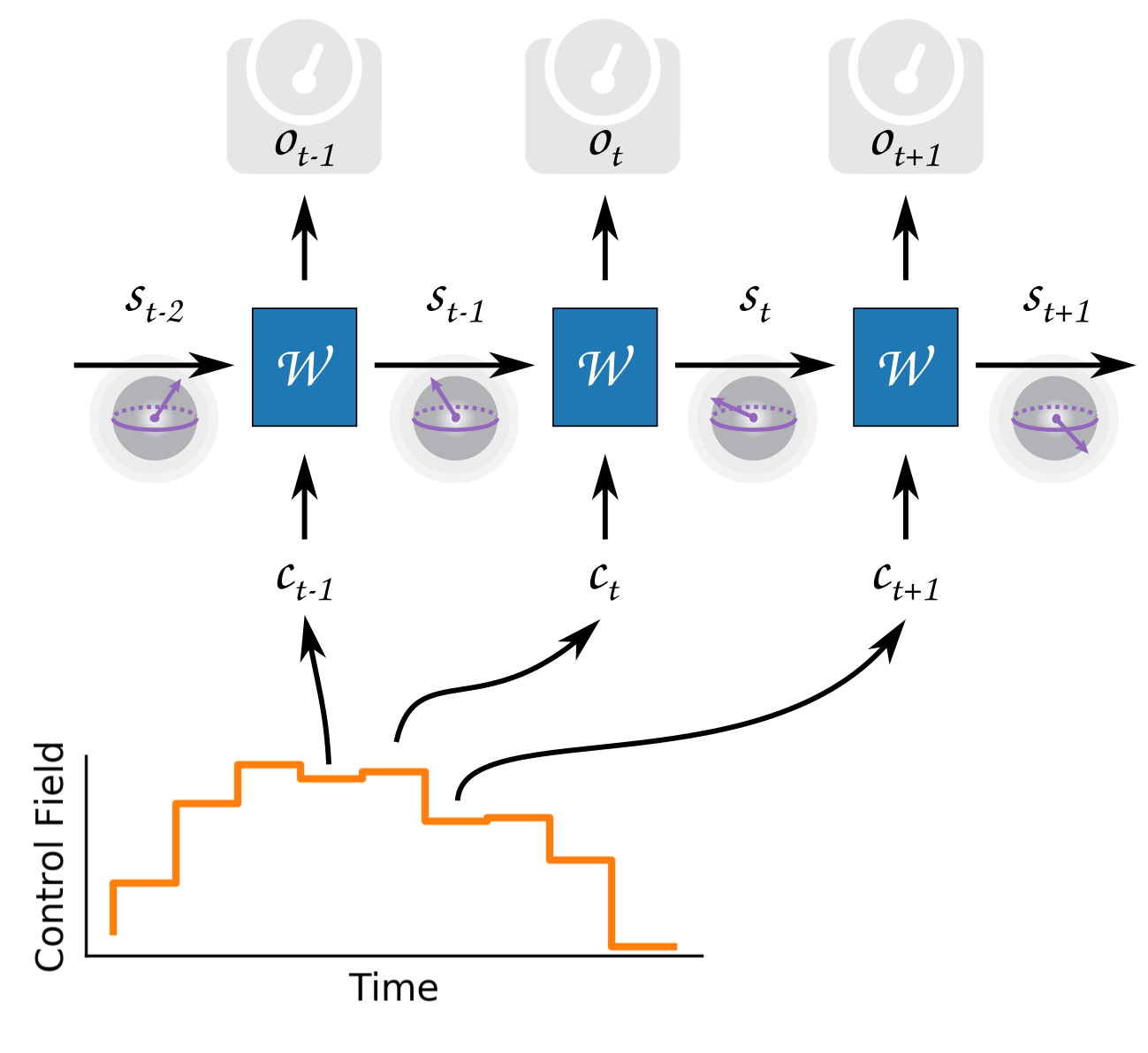}
    \caption{General model of dynamics: An observable $\bm{o}_t$ at time $t$ is interpretable, but the internal state of the system $\bm{s}_t$ (e.g. a Bloch vector or a density matrix) needs not be interpretable outside of the model. A given model is parameterized by a set of values $\mathrm{W}$, which can be an arbitrary parameterization in the case of RNNs, or interpretable interaction rates in a more physical model.}
    \label{fig:rnn}
\end{figure}

While the STEADY approach can be used for optimal control and experimental design, its advantage over the status quo is most clearly elucidated in the task of parameter estimation where we want to learn a model for the dynamics of a system from experimental measurements. To introduce the method, first consider the typical black-box approach popularized by neural networks, or in this case a recurrent neural network (RNN). An RNN explicitly contains the notion of time and instantaneous dynamics. The state of the system is encoded in an uninterpretable and learned vector $\bm{s}_t$ where the subscript denotes the time step, the experimental observations are in the vector $\bm{o}_t$, and control fields imposed on the hardware are in the vector $\bm{c}_t$. The dynamics are encoded in the weights $\mathrm{W}$ which parameterize the behavior of each recurrent step as seen in Fig.\ \ref{fig:rnn}. A typical parameterization where the weights are three matrices $\mathrm{W}=(\mathrm{W_{ss}},\mathrm{W_{sc}},\mathrm{W_{os}})$ might look like

\begin{equation}
\begin{split}
    \bm{s}_t = & \ \mathrm{f}[\mathrm{W},c_t]\circ\mathrm{f}[\mathrm{W},c_{t-1}]\circ\cdots\mathrm{f}[\mathrm{W},c_{0}](\bm{s}_0) \\
    \bm{o}_t = & \ \sigma(\mathrm{W_{os}}\cdot\bm{s}_t),\\
    \mathrm{f}[\mathrm{W},c]:& \ \bm{s} \mapsto \sigma(\mathrm{W_{ss}}\cdot\bm{s}_t+\mathrm{W_{sc}}\cdot\bm{c}),
\end{split}
\end{equation}
where $\sigma$ is some sigmoid function, $\circ$ denotes function application, and $\cdot$ denotes matrix multiplication.

Next we compare this to the classical way to describe the dynamics of a quantum system with a master equation ($\mathrm{ME}$)

\begin{equation}
    \dot{\rho} = \mathrm{ME}(\mathrm{W},\mathrm{c},\rho) = -i\left[\mathrm{H}(\mathrm{W},c),\rho\right],
\end{equation}
where $\mathrm{ME}$ is the given master equations, $\bm{c}$ is the instantaneous value of the control fields imposing the Hamiltonian $\mathrm{H}$, and $\mathrm{W}$ is a set of parameters defining the Hamiltonian (e.g.\ energy levels and coupling rates) and other generators of time evolution. We will explore non-unitary and non-Markovian dynamics below. Unlike the RNN description, this describes only the instantaneous dynamics, but after using an integrator (e.g.\ Euler's method) we can find

\begin{equation}
\begin{split}
    \rho_t = & \ \mathrm{f}[\mathrm{W},c_t]\circ\mathrm{f}[\mathrm{W},c_{t-1}]\cdots\mathrm{f}[\mathrm{W},c_{0}](\rho_0) \\
    o_t = & \ \mathrm{Tr}(\mathrm{O}\rho) \\
    \mathrm{f}[\mathrm{W},c]:& \ \rho \mapsto \rho+\mathrm{ME}(\mathrm{W},\mathrm{c},\rho)\Delta t,
\end{split}
\end{equation}
where $\mathrm{O}$ is the observable we are studying and the time interval $\Delta t$ is the integrator's step size. This is exactly the RNN structure we have already considered (or more precisely, a neural ODE~\citep{chen2018neural}), but with a different parameterization for the dynamics. Advanced adaptive integrators can be used for stiff dynamics without changing this structure.

This identification permits us to work-around the usual constraints of the bias-variance theorem, and also retain the favorable convergence properties of deep neural network models. We can sequester the overparameterized nature of the neural network to the parameterization of the generator of the dynamics, e.g.\ by rewriting the Hamiltonian as neural-net-like expression~\citep{krastanov2019stochastic}:

\begin{equation}\label{param-eq}
\mathrm{H}_{ij}(\mathrm{S},\mathrm{h};\bm{c})=h_{ij}+{\sum}_{k}\mathrm{S}_{ijk}c_{k},
\end{equation}
where $\mathrm{W} = (\mathrm{h}, \mathrm{S})$ are the arrays of parameters that need to be learned. We can employ similar parameterizations for collapse operators and non-Markovian memory kernels.

This parameterization, together with enforcing the overall physics-based form of the master equation, enables high-fidelity parameter estimation resilient to model errors while using orders of magnitude less training data, as our model does not need to learn the laws of nature from scratch. Interestingly, such approaches have recently been suggested in robotics~\citep{lutter2019deep}. Below we explore how to encode various master equations into such a machine learning architecture.

\subsection*{Stochastic Master Equations and Weak Measurements}

\begin{figure}
\includegraphics[width=\columnwidth]{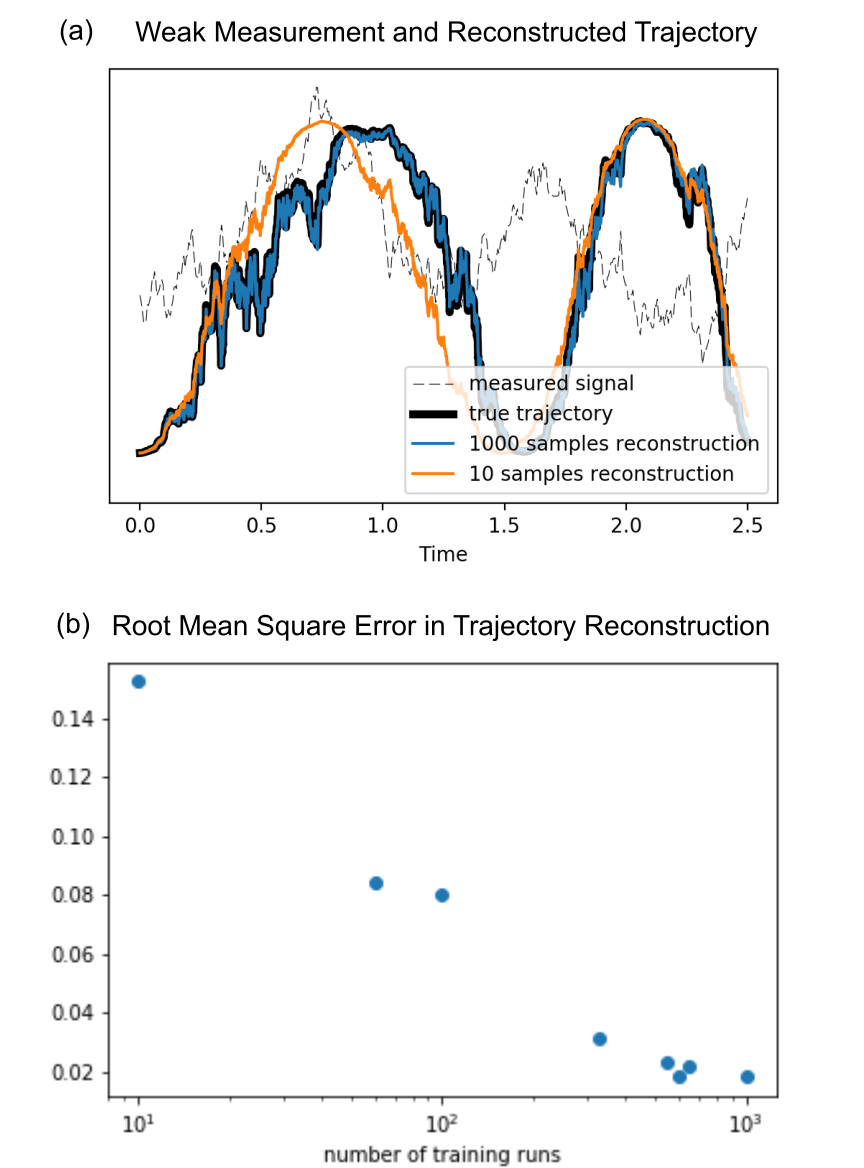}
\caption{Reconstructing the dynamics of a system based on a week measurement record. (a) An illustration of a recorded weak measurement signal and the corresponding reconstruction. The dashed line represents the recorded signal from one single quantum trajectory (e.g., recorded signal from a readout cavity in arbitrary units). The thick black line describes the true, experimentally unavailable (but available to us in the simulation), trajectory that the qubit excited population has taken. The blue and orange lines are our attempts to reconstruct the trajectory from the weak measurements, by using estimates for the dynamical parameters governing the system. For orange the STEADY parameter estimator had access to only 10 recorded trajectories. For blue it had access to 1000 recorded trajectories. The trajectories were recorded at a resolution at least as dense as the one required by the numerical solver. For similar results RNN based methods require millions of recorded trajectories in their training set. (b) A more formal test of the fidelity of reconstruction, plotting the root mean square error in the population prediction, averaged over the entire duration of multiple randomly sampled trajectories.}
\label{fig:example_traj}
\end{figure}

To model the effect of a noisy environment, we need to employ at the very least a master equation like Lindblad's or a quantum trajectories approach~\cite{Lindblad1976, Gorini1976}. However, if we are able to additionally obtain a partial knowledge of what the environment saw when it disturbed our system (i.e., a weak measurement), then we need to extend the Lindblad approach to a more general stochastic master equation.

Performing parameter estimation for stochastic master equations has 
proven particularly difficult, due to the challenges to define an
efficient optimization target over a stochastic process.
Competing schemes capable of dealing with these difficulties require millions of measurement records in their training sets~\citep{flurin2018rnn}, while we achieve similar results with three orders of magnitude fewer measurements, by rewriting a typical backaction master equation solver in an autodifferentiable form.

We consider a system governed by the following dynamics (an Ito stochastic
differential equation)
\begin{equation}
\begin{split}
\mathrm{d}\rho= & \left(-i\left[H,\rho\right]+\mathcal{D}\left(c,\rho\right)\right)\mathrm{d}t+\sqrt{\eta}\mathcal{H}\left(c,\rho\right)dW\\
\mathcal{D}(c,\rho)= & c\rho c^{\dagger}-\frac{1}{2}\left\{ c^{\dagger}c,\rho\right\} \\
\mathcal{H}(c,\rho)= & c\rho+\rho c^{\dagger}-\mathrm{Tr}\left(\rho\left(c+c^{\dagger}\right)\right)\rho. 
\end{split}
\end{equation}
 $H$ is the Hamiltonian governing the unitary evolution of the density
matrix $\rho$. $\mathcal{D}$ marks the Lindblad superoperator with
collapse operator $c$. $\mathcal{H}$ marks the backaction for observing
that same collapse operator, with measurement efficiency $\eta$.
$W$ is a random Wiener process. We use the common notation for commutator
and anticommutator.

The backaction is representing our knowledge of what would otherwise
have been information lost in the dissipative Lindblad dynamics. That
information takes the form of the record of weak measurements $V(t)$.
We can reconstruct the instance of the Wiener process that
has occurred from that record through 
\begin{equation}
\mathrm{d}W=\left(V-2\sqrt{\eta}\mathrm{Tr}\left(\rho c\right)\right)\mathrm{d}t.\label{eq:V}
\end{equation}

To attempt to reconstruct the trajectory of a given state, given this
measurement record we would want to solve the following new Ito stochastic
differential equation, derived from the original master equation after
substituting in the noise reconstruction procedure: 
\begin{multline}
\mathrm{d}\rho=\big(-i\left[H,\rho\right]+\mathcal{D}\left(c,\rho\right)-2\eta\mathcal{H}(c,\rho)\mathrm{Tr}\left(\rho c\right)\big)\mathrm{d}t\\+\sqrt{\eta}\mathcal{H}(c,\rho)dV.\label{eq:SDErec}
\end{multline}

We will present the performance of our "stochastic estimator for stochastic master equations" by running it on a simulated hardware, so that we can validate the result against the actual state of the system. We compare our results to other methods on the following system: a single qubit coupled to a readout cavity with $H=\Omega\sigma_{x}$
and $c=\sqrt{\gamma}\sigma_{z}$. The parameters $\Omega$, $\gamma$,
and $\eta$ are to be estimated from the measurement record. For each
trajectory we record a sequence of measurements
$\{V_{t_{0}},V_{t_{1}},\dots\}$ and a final projective measurement
on the z basis (one single bit of information). We train our auto-differentiable stochastic master equation model on this data. We evaluate the fidelity of the
estimates by trying to reconstruct a separate set of trajectories based on their weak measurement record (a validation data set). An example of attempting to reconstruct one such trajectory can be seen in Fig.~\ref{fig:example_traj}.

This approach can be viewed as restating a stochastic differential equation solver in the form of a recurrent machine learning model. The explicit structure we add to that model, i.e.\ the form of the stochastic differential equation, ensures we do not need to re-learn universal truths of physics (like the CPTP properties of quantum dynamics), while keeping the results interpretable, and requiring drastically lower amounts of training data. At the same time, a richly parameterized generator of the dynamics (i.e., a more general parameterization for $H$ and $c$ if necessary) would allow the same versatility as more naive machine learning models, without requiring an excessively large training set.

\subsection*{Non-Markovian Environments}

\begin{figure}
    \centering
    \includegraphics[width=\columnwidth]{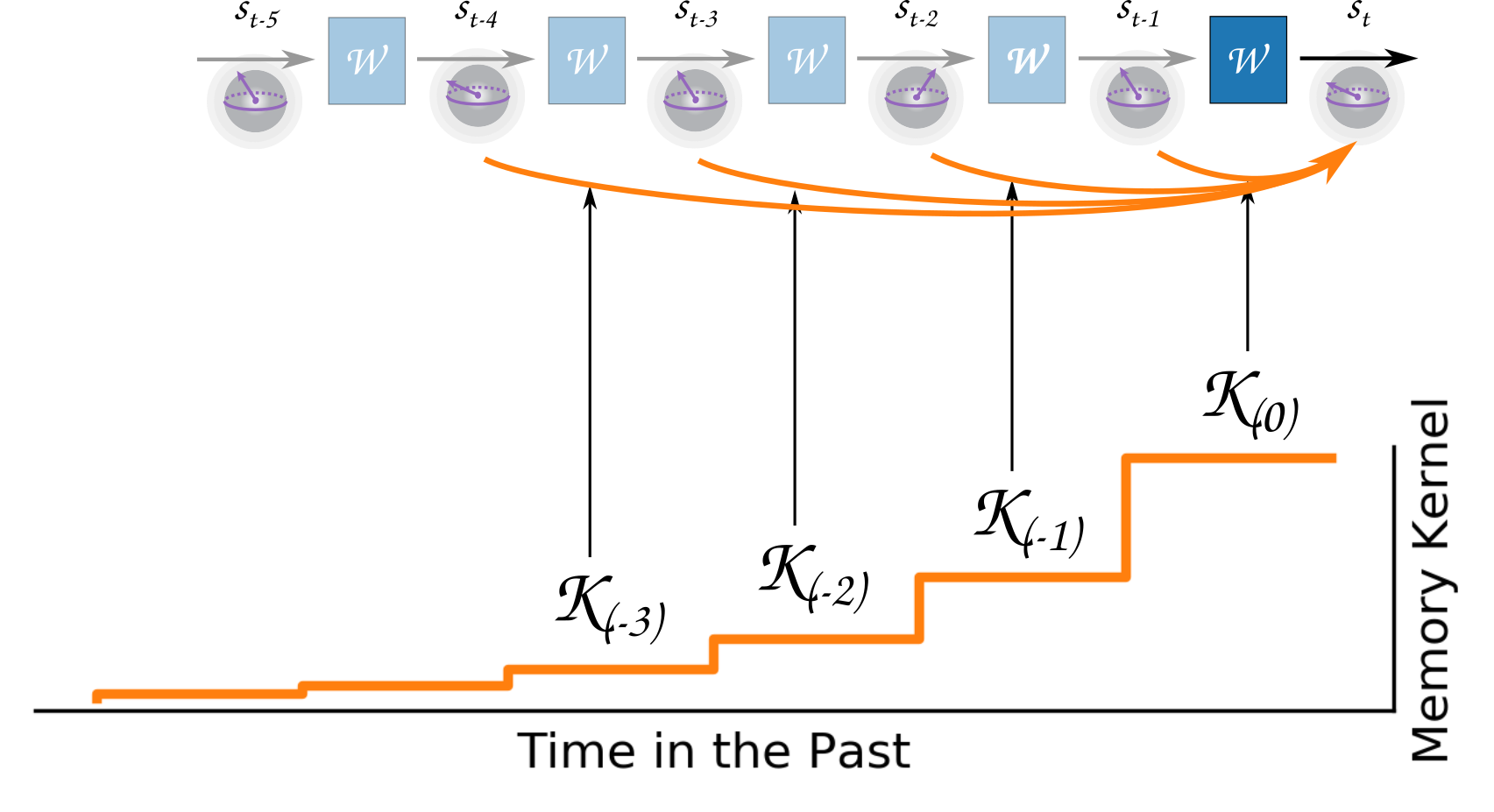}
    \caption{Introducing a memory kernel, as done in the Nakajima-Zwanzig equation is equivalent to introducing a set of "shortcuts" in our pseudo-RNN interpretation of master equations. These shortcuts implement a "memory" by enabling old states to affect the current dynamics. The memory kernel, which has to be learnt, encodes the weight of each shortcut as a function of its time-horizon.}
    \label{fig:rnn_nz}
\end{figure}

The STEADY methodology of recasting master equations into machine learning models with explicitly physical architecture can be extended further to one of the most general types of quantum dynamics. The Nakajima–Zwanzig formalism is one of the most accurate methods of generating completely positive trace-preserving dynamics, free of the various assumptions usually made about the dynamics governing the environment~\cite{Nakajima1958, Zwanzig1960, Breuer2002}. This method removes the assumption that the bath is Markovian by introducing the memory kernel $\mathcal{K}$,
\begin{equation}
\begin{split}
\frac{\mathrm{d}\rho}{\mathrm{d}t}= & -i\left[H,\rho(t)\right]+\int_0^t\mathcal{K}(\tau)\mathcal{D}\left(c,\rho(t-\tau)\right)\mathrm{d}\tau\\
\end{split}
\end{equation}
The above form assumes of an initially separable system-bath state~\citep{Breuer2002}.
This generalized master equation is often the starting point for other theories of quantum evolution, for example Lindbladian dynamics can be recovered when $\mathcal{K}$ becomes the Dirac $\delta$ distribution~\cite{Breuer2002}. This parameterization can be derived from first principles if the goal is to describe classical  non-Markovian noise. Moreover, as an ansatz, it is a fruitful way to describe coherent effects in the (otherwise instantaneously decohering) baths. Coupled with our parameterization and optimal control framework this formalism enables using the non-Markovian environment as a control resource~\citep{Li2020}. Possible applications include improving the coherence of qubits surrounded by parasitic two-level systems whose effects are usually modeled as simple decoherence or controlling systems subjected to colored noise. This modeling problem becomes increasingly important as we attempt high-fidelity control of superconducting 

Figure\ \ref{fig:rnn_nz} describes the main difficulty in implementing this approach. Up to now we were able to restate differential equations in the form of a particularly structured recurrent model, but the memory kernel involves an integral-differential equation which introduces "shortcuts" between the layers of the model. Implementing it in a form that is amenable to automatic differentiation creates additional time complexity, proportional to the depth of the memory kernel.

\section*{Discussion}

\begin{figure}
    \centering
    \includegraphics[width=\columnwidth]{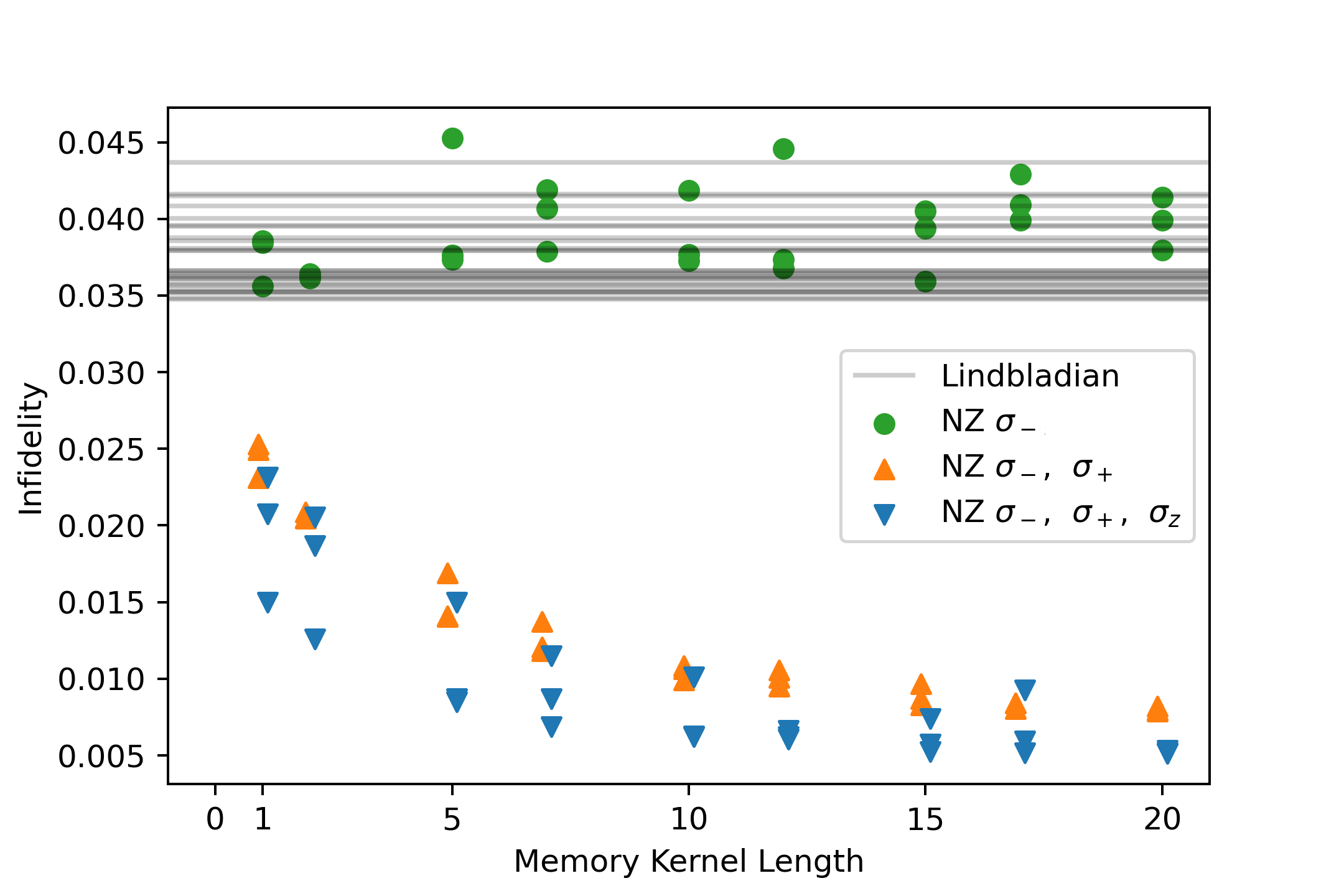}
    \caption{The predictive power of a trained Nakajima-Zwanzig model. We prepare training data by performing a complete unitary simulation of a qubit subjected to randomized control drives and coupled to a parasitic system (a spin-star bath). A memory kernel is trained on that data after the bath has been traced out. The abscissa shows the length of the trained kernel in $\Delta\tau$ units, vs the infidelity of the obtained predictions on the ordinate. The gray lines are the predictions from a purely Lindbladian model. In green we have a NZ model with only a $\sigma_-$ collapse operator in the memory kernel, which, similarly to the Lindbladian model, proves to be insufficient. The orange and blue markers show the performance of memory kernels that also contain $\sigma_+$ excitation operator and $\sigma_z$ dephasing operator. Each marker is based on gradient descent over synthetic data, with only 20 sample trajectories in the training set. A small swarm of independent optimizations was employed for each model, hence the multiple markers of the same shape. The minimum in each group represents the best model, while the spread is indicative of the numerical difficulty of the optimization problem. The infidelity is the root mean square error in the model prediction for the qubit excited population.}
    \label{fig:nz_comparison}
\end{figure}

To test the feasibility of our auto-differentiated approach to directly train the Nakajima-Zwanzig equation we picked a system consisting of a qubit surrounded by parasitic two-level systems coupled to it~\cite{Breuer2004, Breuer2007}. Such a bath causes more than decoherence, and a simple Lindbladian model that traces out the parasitic subsystems does not provide a high-fidelity reconstruction of the dynamics. At the same time, that type of interactions become increasingly important in accounting for the imperfections of superconducting and NV qubits. A direct unitary simulation of these dynamics requires knowledge of the number of parasitic subsystems and learning their complex network of couplings. Moreover, such a simulation would be exponentially expensive in the number of parasites. Deep learning methodologies would certainly be useful here~\citep{Banchi2018}, but they have all of the already mentioned downsides pertaining to their significant training data requirements. On the other hand, this system is a perfect target for a Nakajima-Zwanzig approach with a richly parameterized kernel. We overparameterize only the generator of the dynamics, in this case, the memory kernel, while we base the overall architecture on a well defended general master equation. In Fig.\ \ref{fig:nz_comparison} we see how this method makes it possible to learn the dynamics to an otherwise unreachable fidelity.

Our approach enables us to restate known master equations in a way that is amenable to the optimization techniques in machine learning. By choosing the part of the model with respect to which we perform the gradient descent, this approach can be used simultaneously for parameter estimation, optimal control, or experiment design. In particular, the precise characterization of non-Markovian baths enabled by this augmented framework opens the door for noise-aware control of devices, where coherences in the interaction with the environment can be used as a resource --- one which is frequently disregarded. Here, we exemplify this method on a spin star system where the parasitic couplings to slow baths have been revealed to be the main outstanding systematic error in the modeling of cutting edge quantum hardware~\citep{niu2019learning}.

The non-Markovian STEADY paradigm, provides an important path forward, elusive to other techniques. We retain the rich parameterization that facilitates generalizability and avoidance of model errors, seen in ML methods. However, our protocol does not need vast amounts of training data of deep learning, as the universal properties of the generator of the physical dynamics is explicitly encoded in the structure of our `pseudo-neural network'. Our versatile physics-based modeling approach would make possible the high-fidelity calibration and noise-aware control of quantum systems even in the presence of non-Markovian effects.



\section*{Methods}

\subsection*{Models and computational framework}

Neural networks have been lauded for their "universal approximator" properties, however, as already alluded to through the mention of the bias-variance theorem, this universality comes at a significant cost in the amount of necessary training data. Here we present more explicitly the software stack we are using for our physically-constrained universal approximators, enabled by the engineering work that the machine learning community has made freely available.

We used differentiable-programming frameworks (Tensorflow~\citep{diffprog_tensorflow2015-whitepaper} and Zygote~\citep{zygote}) in order to create Schrodinger and Lindblad equation solvers, as well as solvers for the stochastic master equation and Nakajima-Zwanzig integral-differential equation. The solver types included Euler, Euler–Maruyama, and Runge–Kutta methods. Simplified and cleaned up version of these solvers is provided as supplementary files. The aforementioned autodifferentiation frameworks permit the efficient calculation of cost functions involving the solutions of these equations.

The parameterization of the generators of the dynamics is left unconstrained. In simple cases the parameterization can remain very sparse, but when necessary the various couplings themselves can be represented as the output of a single-layer (or deeper) neural net, similar to Eq.\ \ref{param-eq}. A particularly interesting example is the system we introduced when discussing memory kernels: a qubit surrounded by an unspecified number of parasitic two-level systems of various couplings. The unitary model of such a system would involve a large number of explicitly parameterized couplings in a relatively large Hilbert space. On the other hand, the Nakajima-Zwanzig model hides all of this complexity in a small number of Lindblad-like terms, each having a parameterized memory-kernel profile $\mathcal{K}(\tau)$ that needs to be learnt.

\subsection*{Choice of test systems and cost functions for non-Markovian dynamics}

\begin{figure}
    \centering
    \includegraphics[width = 0.6\columnwidth]{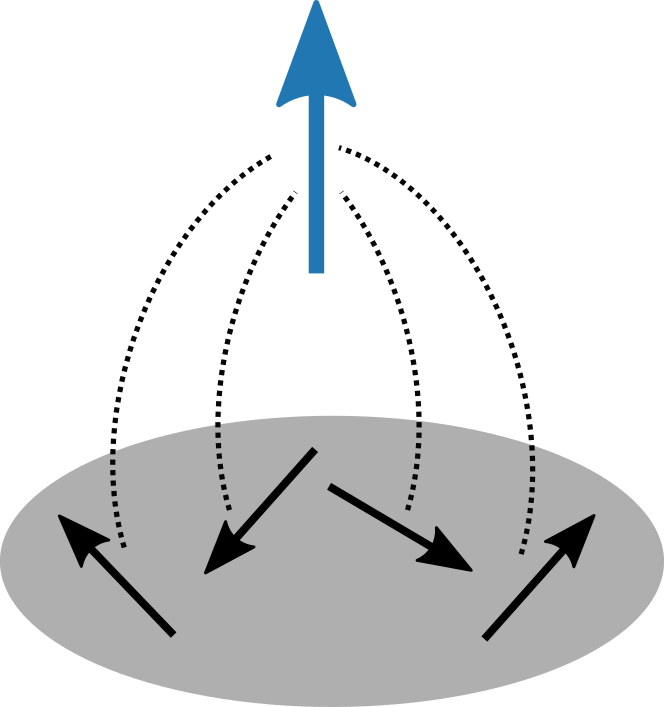}
    \caption{Central system spin interacting with and environment of parasitic two-level systems. Encountered when attempting to control superconducting qubits or NV centers, due to interactions with their surroundings. Modeling such systems directly, is costly, as it involves the simulation of multiple two-level subsystems. Tracing out the parasitic systems leads to dynamics much more complicated than a typical Lindbladian master equation, because the characteristic timescale of the bath is not much faster than that of the central qubit. This leads to the need for the Nakajima-Zwanzig formalism, employing integral-differential equations with memory kernels. In a way, parameter estimation over such an ansatz can be simpler, because we do not need to worry about the exact number of parasitic two-level systems and their exact couplings. }
    \label{fig:spinstar}
\end{figure}

The spin star system is a good test case of a qubit interacting with a parasitic spin bath~\citep{Breuer2004}. It has shown to be an effective approximate method for modeling the dynamics of many solid-state spin systems, including nitrogen-vacancy centers in diamond~\citep{Zhao2012} and semiconductor qubits~\citep{Radhakrishnan2019}, where the electronic system of interest interacts with a nuclear spin environment. To produce mock training data for the STEADY method, the entire five spin system depicted in Fig.~\ref{fig:spinstar} was simulated using the full unitary dynamics governed by the complete system-bath Hamiltonian as given by,

\begin{align}
   \hat{H} = \frac{\omega_0}{2}\sigma_z + \sum_{k=1}^{N} A_k \sigma_x \sigma_x^{(k)} + \epsilon_x(t)\sigma_x + \epsilon_y(t)\sigma_y,
\end{align}
where $\omega_0$ is the system energy gap, $\sigma_\square$ and $\sigma_\square^{(k)}$ are the Pauli operators for the system and $k$th environmental spin respectively (subscript denoting $x$, $y$, or $z$), and $\epsilon_\square(t)$ is a time-dependent driving field (along $x$ or $y$, in subscript).  

Instead of training this entire model, which can include a large number of arbitrarily coupled two-level systems, we train the NZ model which uses a much simpler parameterization. Moreover, the exponentially large cost of simulating multiple subsystems is avoided, as we are simulating one single two-level system. In its place, we have the increased computational complexity due to the presence of a time kernel, proportional to the length of that very kernel. Practically, that length can be expected to scale as the slowest dynamics involving the traced-out bath, however the exact time and space complexity required of the NZ equation are still an open question. Nonetheless, the NZ modeling approach provides a hyperparameter enabling an otherwise inaccessible trade-off: a shorter kernel can be used when the need for fast computations outweighs the need for high fidelity (as already seen in Fig.\ \ref{fig:nz_comparison}).

The training process itself involves minimizing the distance between the model prediction and the training data. In the case of the weak-measurement dynamics, it is impossible to repeat the same trajectory multiple times, hence the cost function is

\begin{align}
   \hat{C} = \frac{1}{S} \sum_{i=1}^{S} \mathrm{d}\left(p_i^{(e)}, p_i^{(m)}\right),
\end{align}
where $S$ is the number of recorded sample trajectories (indexed by $i$), $p_i^{(m)}$ is the Born probability for the qubit based on the current model parameters, and $p_i^{(e)}$ is the corresponding experimental measurement (which can only have the values 0 or 1 given that we can sample that trajectory only once). $\mathrm{d}$ is some distance function, e.g. the square of the difference.

If no weak measurements are performed (e.g., the test system we used for the NZ model), then the same trajectory can be sampled repeatedly, and at multiple different times. One possible cost function in that case would be

\begin{align}
   \hat{C} = \frac{1}{S} \frac{1}{T} \sum_{i=1}^{S} \sum_{j=1}^{T} \mathrm{d}\left(p_i^{(e)}(t_j), p_i^{(m)}(t_j)\right),
\end{align}
where $T$ denotes the number of different durations at which a given experimental replicate is sampled (indexed by $j$ and $t_j$ is the corresponding time). Both $p_i^{(e)}$ and $p_i^{(m)}$ are now evaluated at multiple different times. Moreover $p_i^{(e)}$ can now be sampled repeatedly for higher resolution of the training data.

\begin{acknowledgments}

\noindent We thank the Python, Numpy, Tensorflow, and Julia opensource communities, as well as Yale's HPC and Harvard's RC teams, for facilitating the numerical work involved in this project. K.H.M. and S.K. are partially supported by the U.S. Department of Energy, Office of Science, Basic Energy Sciences (BES), Materials Sciences and Engineering Division under FWP ERKCK47 `Understanding and Controlling Entangled and Correlated Quantum States in Confined Solid-state Systems Created via Atomic Scale Manipulation'.
This work is also supported by the Harvard Physical Sciences Accelerator Award and Harvard Quantum Initiative Seed Grant. This work was completed while STF was visiting the Yale Quantum Institute, and he is grateful for their hospitality. S.Z. and L.J. acknowledge support from the National Science Foundation (OMA-1936118) and the Packard Foundation (2013-39273).
P.N. is a Moore Inventor Fellow and gratefully acknowledges support through Grant GBMF8048 from the Gordon and Betty Moore Foundation.
\end{acknowledgments}

\bibliography{references}

\end{document}